\documentclass[aps,twocolumn,superscriptaddress,longbibliography]{revtex4-2}
\usepackage[colorlinks=true, citecolor=blue, urlcolor=blue, linkcolor=red]{hyperref}
\usepackage{amssymb,physics,orcidlink}

\begin{document}
\title{Noise-resilient quantum reinforcement learning}
\author{Jing-Ci Yue\orcidlink{0009-0002-0934-3192}}
\affiliation{Key Laboratory of Quantum Theory and Applications of MoE, Lanzhou Center for Theoretical Physics, Gansu Provincial Research Center for Basic Disciplines of Quantum Physics, and Key Laboratory of Theoretical Physics of Gansu Province, Lanzhou University, Lanzhou 730000, China}
\author{Jun-Hong An\orcidlink{0000-0002-3475-0729}}\email{anjhong@lzu.edu.cn}
\affiliation{Key Laboratory of Quantum Theory and Applications of MoE, Lanzhou Center for Theoretical Physics, Gansu Provincial Research Center for Basic Disciplines of Quantum Physics, and Key Laboratory of Theoretical Physics of Gansu Province, Lanzhou University, Lanzhou 730000, China}

\begin{abstract}
As a branch of quantum machine learning, quantum reinforcement learning (QRL) aims to solve complex sequential decision-making problems more efficiently and effectively than its classical counterpart by exploiting quantum resources. However, in the noisy intermediate-scale quantum (NISQ) era, its realization is challenged by the ubiquitous noise-induced decoherence. Here, we propose a noise-resilient QRL scheme for a quantum eigensolver with a two-level system as an agent. By investigating the non-Markovian decoherence effect on the QRL for solving the eigenstates of the agent-environment interaction Hamiltonian, we find that the formation of a bound state in the energy spectrum of the total agent-noise system restores the QRL performance to that in the noiseless case. Providing a universal physical mechanism to suppress the decoherence effect on quantum machine learning, our result lays the foundation for designing NISQ algorithms and offers a guideline for their practical implementation.
\end{abstract}
\maketitle

\section{Introduction}
The rapid development of quantum technologies and artificial intelligence raises an important question: What happens if we combine machine learning and quantum physics? The answer is an algorithm revolution. Running algorithms on quantum devices enables quantum machine learning to solve data processing, pattern recognition, and optimization problems more powerfully than its classical counterpart by exploiting quantum effects \cite{RN41,RN32,PRXQuantum.2.040321}. One primary type of quantum machine learning is quantum reinforcement learning (QRL). It distinguishes itself from others, such as quantum supervised \cite{Innocenti_2020,RN3,RN30,PhysRevA.99.042327,PhysRevLett.122.040504,PhysRevA.98.032309,PhysRevA.101.032308,PerezSalinas2020datareuploading} and unsupervised \cite{PhysRevA.94.022308,RN50,PhysRevX.8.021050,RN52,doi:10.1126/science.abn7293,PhysRevLett.126.240402,RN54} learning, in being able to efficiently find an optimal policy via evaluating the mapping from states to actions without labeled data \cite{4579244,PhysRevLett.117.130501,PhysRevA.98.042315,PhysRevX.4.031002,https://doi.org/10.1002/qute.201800074,Moro2021}. It can solve the difficulty of classical situation wherein the action-space dimension grows exponentially with the complexity of the task \cite{4579244}, which is dubbed the curse of dimensionality. QRL, in turn, promotes the development of quantum technologies. Its significant capabilities have been demonstrated in quantum control \cite{PhysRevX.8.031086,PhysRevLett.126.060401,PhysRevA.103.012404}, state transfer \cite{PhysRevA.97.052333}, communication \cite{PRXQuantum.1.010301}, and sensing \cite{RN5,Schuff_2020,PhysRevLett.134.120803}.

\begin{figure}[tbp]
\centering
\includegraphics[width=\columnwidth]{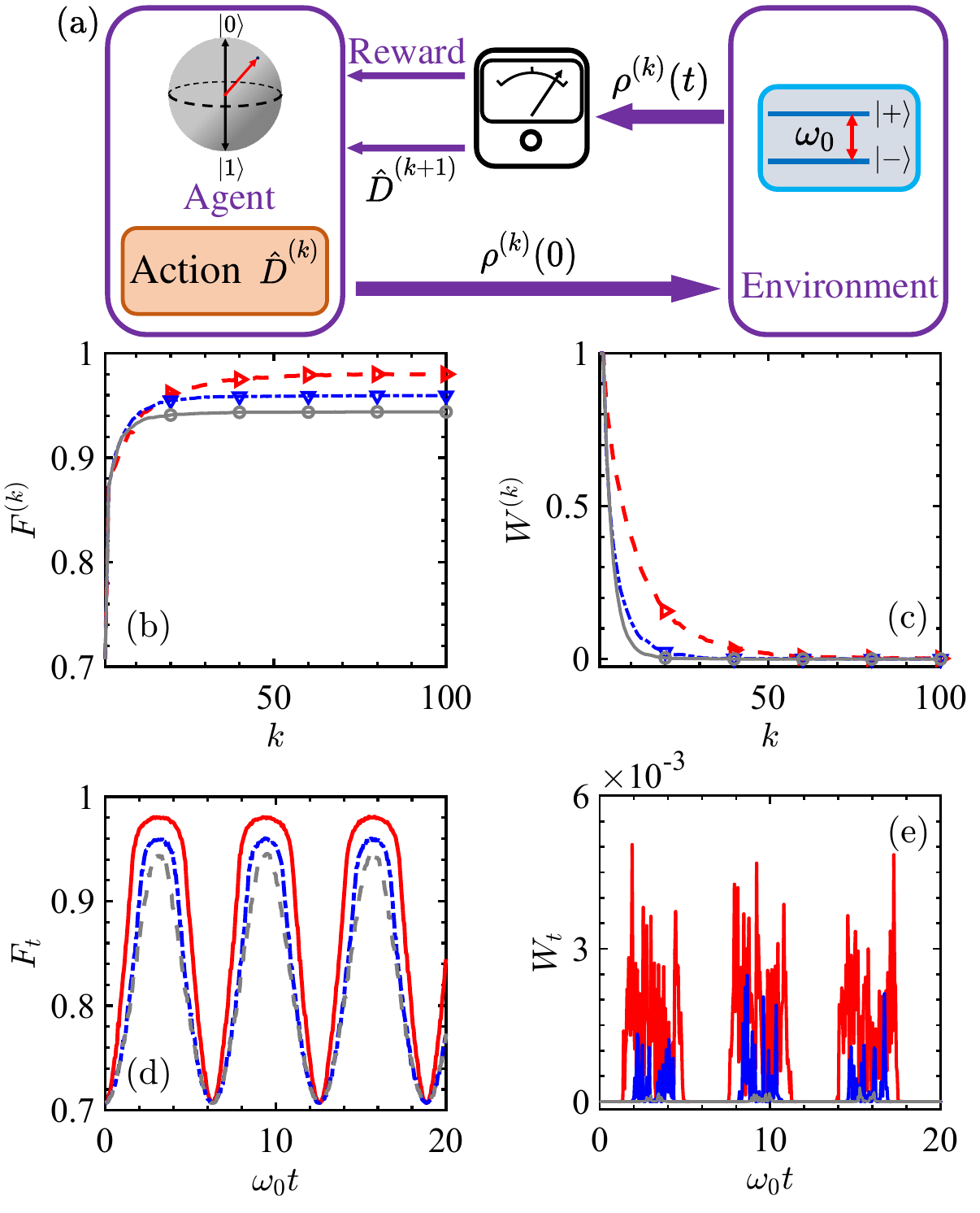}
\caption{(a) QRL protocol. (b) Mean fidelity and (c) mean exploration parameter as functions of the iteration times $k$ for different values of $(r,p)$ for $ t=15.65\omega_0^{-1}$. (d) Mean fidelity and (e) mean exploration parameter as a function of the interaction time $t$ in different values of $(r,p)$ for $k=100$. $(r,p)=(0.5,2)$, $(0.1,2)$ and $(0.1,1.1)$ appear as red, blue, and gray lines in (b-e). We use $N=1000$.}\label{fig1}
\end{figure}

Despite all this substantial progress, the practical implementation of quantum algorithms on quantum devices is still challenged by the decoherence caused by different kinds of noise in the quantum world. This is an important feature of the noisy intermediate-scale quantum (NISQ) era \cite{RevModPhys.94.015004,Preskill2018quantumcomputingin,PhysRevA.105.052445}. The inherent fragility of quantum effects has been shown to pose a severe obstacle to achieving the full potential of quantum technologies \cite{PhysRevA.100.043833,PhysRevLett.122.210601,PhysRevX.12.011039,PhysRevLett.129.240503,PhysRevLett.106.020501}. In terms of quantum algorithms, noise-induced decoherence also degrades the performance of various quantum algorithms \cite{PhysRevA.65.042311,Marconi2022roleofcoherence,PhysRevA.93.012122,RN39,PhysRevApplied.14.014100,PhysRevA.91.062320,PhysRevResearch.5.033159,PhysRevE.108.014128,Sun2024,Deng2013}. Typical examples include: the noise-induced barren plateau phenomenon \cite{RN39}, where decoherence causes the gradients of the objective function in variational quantum algorithms to vanish exponentially with increasing system size; and the deformation of the ground state in adiabatic quantum computation \cite{Deng2013}, where decoherence induces a loss of fidelity of the ground state and failure of the calculation. Decoherence also causes the degradation of the fidelity of QRL \cite{PhysRevE.108.014128}.

Many strategies have been designed to overcome the detrimental effects of decoherence on NISQ algorithms, including the hybrid quantum-classical approach \cite{RN37}, dynamical decoupling \cite{van2012}, quantum error mitigation \cite{PhysRevLett.119.180509,PhysRevA.100.010302,PhysRevA.95.042308} and correction \cite{RN40}, adding ancillas \cite{Liao2023}, NISQ reservoir computing \cite{Hu2024}, and quantum neural networks \cite{RN7}. However, those works are exclusively based on the Born-Markov approximation to describe the decoherence. Given the essential differences of non-Markovian dynamics from the Born-Markov approximate one \cite{RevModPhys.88.021002}, this description is obviously insufficient to reveal the intrinsic robustness of quantum algorithms to decoherence. It has been found that the non-Markovian effect can be harnessed as a resource to enhance the resistance of different quantum technologies against decoherence \cite{PhysRevA.102.012217,PhysRevLett.124.140502,PhysRevLett.123.040402}. Hence, how to develop a noise-resistant QRL scheme under non-Markovian decoherence remains an open question.

Here, via investigating the noise-induced decoherence on the two-level system as the agent of the QRL, we propose a noise-resilient QRL scheme for a quantum eigensolver. In sharp contrast to one's general belief that the performance of the QRL is deteriorated by the decoherence under the Born-Markov approximation, we discover a physical mechanism in preserving the ideal performance of the QRL in the exact non-Markovian dynamics. Our analysis reveals that it is due to the formation of a bound state in the energy spectrum of the total system consisting of the agent and the noise. Supplying a useful method for overcoming the destructive impact of decoherence on the QRL, our result paves the way for its practical realization and application.

This paper is organized as follows. In Sec. \ref{sec:noiseless}, we present the noiseless QRL protocol. In Sec. \ref{sec:noise}, we investigate the influence of the non-Markovian decoherence of the dissipative noise on the QRL protocol. The significant enhancement of the mean fidelity reveals the vital role played by the agent-noise bound state. Section \ref{sec:realization} displays a physical platform constructed in the circuit-QED system to realize our noise-resilient scheme. We draw conclusions in  Sec. \ref{sec:conclusion}.

\section{Noiseless QUANTUM REINFORCEMENT LEARNING}\label{sec:noiseless}
To uncover the effect of the noise on the QRL protocol, we select a minimal QRL scheme  for a quantum eigensolver \cite{Albarrán-Arriagada_2020}.
It comprises an agent and an environment [see Fig. \ref{fig1}(a)]. The agent is a controllable quantum system in an input state $|\psi(0)\rangle$. The environment is a classical system and acts as a black box that interacts with the agent for a duration of time $t$. The interaction is governed by an evolution operator $\hat{U}_t=e^{-i\hat{H}t/\hbar}$ acting on $|\psi(0)\rangle$, where $\hat{H}$ is an unknown Hamiltonian whose eigenstates are underdetermined \cite{Albarrán-Arriagada_2020,PhysRevE.108.014128}. For explicitness, we restrict our analysis to a two-level system with the bare basis $\{|0\rangle,|1\rangle\}$ and ($\hbar=1$),
\begin{equation}
\hat{H}=\frac{\omega_0}{2}(\ket{+}\bra{+}-\ket{-}\bra{-}),
\end{equation}
where $\omega_0$ is a constant with dimension of frequency and $\ket{\pm }=(\ket{0}\pm\ket{1})/\sqrt{2}$ are underdetermined eigenstates. Many iterations are executed so that the agent state at the beginning of the $k$th iteration reads $|\psi^{(k)}(0)\rangle$, with $k\in\mathbb{Z}$. The agent in the first iteration is in $|\psi^{(1)}(0)\rangle=|0\rangle$, which relates to $|\psi^{(k)}(0)\rangle$ via the so-called action $\hat{D}^{(k)}$ as $|\psi^{(k)}(0)\rangle=\hat{D}^{(k)}\ket{0}$, with $\hat{D}^{(1)}=I$. The goal of the QRL is to approach the eigenstates $|\pm\rangle$ iteratively by performing a sequence of actions $\hat{D}^{(k)}$ on the agent state. Note that the QRL is applicable to the eigensolver for the multiple-qubit and high-dimensional situation  \cite{Albarrán-Arriagada_2020}. The construction of $\hat{D}^{(k+1)}$ from $\hat{D}^{(k)}$ is as follows.

In the first step of the $k$th iteration, the agent interacts with the environment for a time duration $t$ and its state becomes $|\psi^{(k)}(t)\rangle=\hat{U}_t|\psi^{(k)}(0)\rangle$. In the second step, a measurement of $\hat{M}^{(k)}=\hat{D}^{(k)}|1\rangle\langle 1|\hat{D}^{(k)\dag}$ is made. After obtaining the result $m^{(k)}$ with probabilities $P^{(k)}_m$, the state collapses to $|m^{(k)}\rangle$, with $m^{(k)}=0$ or 1. To numerically simulate the process, a pseudorandom number $\chi^{(k)}$ uniformly distributed in the interval $[0, 1]$ is drawn. If $\chi^{(k)}\le P^{(k)}_m$, then the measurement outcome is $m^{(k)} = 0$. If $\chi^{(k)}> P^{(k)}_m$, then $m^{(k)} = 1$. In the third step, a pseudorandom rotation
\begin{equation}
\hat{R}^{(k)}=e^{-i\alpha^{(k)}_{y}\hat{\sigma}^{(k)}_{y}/2}e^{i\alpha^{(k)}_{z}\hat{\sigma}^{(k)}_{z}/2}e^{-i\alpha^{(k)}_{x}\hat{\sigma}^{(k)}_{x}/2}
\end{equation}
is performed on the agent, where $\hat{\sigma}^{(k)}_{\nu}=\hat{D}^{(k)}\hat{\sigma}_{\nu}\hat{D}^{(k)\dagger}$ and $\hat{\sigma}_{\nu}$ ($\nu=x, y, z$) are the Pauli matrices defined in the bare basis.

The angles $\alpha_{\nu}^{(k)}$ uniformly distributed in the exploration interval $[-w^{(k)}\pi,w^{(k)}\pi]$, where $w^{(k)}$, called the exploration parameter, is computed inductively from $w^{(1)} = 1$, are conditioned by the measurement result $m^{(k)}$. The reward function is defined by the relation between $w^{(k+1)}$ and $w^{(k)}$ as $w^{(k+1)}=\min\small \{ 1,[(1-m^{(k)})r+m^{(k)}p]w^{(k)} \small \}$, where $r\in(0,1)$ is the reward rate and $p>1$ is the punishment rate \cite{Albarrán-Arriagada_2020}. If $m^{(k)} = 1$, then a punishment is applied by increasing $w^{(k)}$ to $w^{(k+1)}=\min[1,pw^{(k)}]$, thus widening the exploration interval. If $m^{(k)} = 0$, then a reward is granted by decreasing $w^{(k)}$ to $w^{(k+1)}=rw^{(k)}$, thus narrowing the exploration interval. When $w^{(k)}$ converges to zero after multiple iterations, the protocol is valid. Finally, $\hat{D}^{(k+1)}$ is constructed from $\hat{D}^{(k)}$ as
\begin{equation}
\hat{D}^{(k+1)}=\big[(1-m^{(k)})I+m^{(k)}\hat{R}^{(k)}\big]\hat{D}^{(k)}.\label{D(k)}
\end{equation}
Therefore, the trade-off between exploration and exploitation, which is a characteristic of reinforcement
learning, is controlled by the measurement outcome $m^{(k)}$. If $m^{(k)}=1$, the agent decides to explore and makes the rotation $\hat{R}^{(k)}$. On the contrary, if $m^{(k)}=0$, the agent decides to exploit and keeps its action invariant.

The performance of the QRL is quantified by the fidelity between the state $|\psi^{(k)}(0)\rangle$ and the closest eigenstate of $\hat{H}$ for each iteration. Since there is no prior knowledge on whether the underdetermined eigenstate is $|+\rangle$ or $|-\rangle$, we take the greater of the two values, i.e.,
\begin{equation}
       f^{(k)}=\max[\,|\langle +|\psi^{(k)}(0)\rangle|, |\langle -|\psi^{(k)}(0)\rangle|\,].
  \end{equation}
The closer the value of $f^{(k)}$ is to $1$ as $k$ increases, the more accurately the eigenstate would be obtained. The convergence extent of each iteration is quantified by the exploration parameter $w^{(k)}$. The faster it approaches zero, the faster the protocol converges.

In the numerical simulations, the protocol with given interaction time duration $t$ and iteration times $k$ is repeated a large number $N$ of times. Thus, the mean fidelity and the mean exploration parameter are $F^{(k)}=(1/N)\sum^{N}_{j=1} f_{j}^{(k)}$ and $W^{(k)}=(1/N)\sum^{N}_{j=1} w_{j}^{(k)}$, where $j$ labels the $j$th execution of the protocol with $k$ iterations. Figures \ref{fig1}(b) and \ref{fig1}(c) show the calculated $F^{(k)}$ as a function of the iteration times $k$ with different values of $(r,p)$ for a given interaction time $t$. With increasing $k$, $F^{(k)}$ increases to a stable value and $W^{(k)}$ tends to $0$. The larger either $r$ or $p$ becomes, the closer $F^{(k)}$ is to 1, while the slower $W^{(k)}$ tends to 0. We have checked that, when $(r,p)=(0.9,20/9)$, the mean fidelity can reach 0.995, which reaches the performance benchmark in single-qubit compiling tasks in simulation with well-tuned learning protocols \cite{Moro2021}. It indicates that a higher fidelity always needs more iteration times. We denote the mean fidelity and the
mean exploration parameter for a given large $k$ as $F_t$ and $W_t$; here $F_t$ exhibits a periodic oscillation with $t$ and $W_t$ remains zero [see Fig. \ref{fig1}(d)]. Therefore, choosing the proper interaction time is a prerequisite for the QRL.

\section{Effect of quantum noise}\label{sec:noise}
The agent-environment interaction is influenced by ubiquitous decoherence \cite{doi:10.1126/science.aax3766,PhysRevA.100.043833,RN57}, which is caused by the interaction between the agent and quantum noise. The Hamiltonian of the total system composed of the agent and the noise is
\begin{equation}
\hat{H}_{T}=\hat{H}+\sum_{n}\left[\omega_{n}\hat{a}^{\dagger}_{n}\hat{a}_{n}+g_{n}(\hat{\sigma}_{-}\hat{a}^{\dagger}_n+\text{H.c.})\right],\label{tthmt}
\end{equation}
where $\hat{a}_{n}$ is the annihilation operator of the $n$th mode with frequency $\omega_{n}$ of the noise and $\hat{\sigma}_-=\ket{-}\bra{+}$ is the transition operator from $|+\rangle$ to $|-\rangle$. The coupling strength  $g_{n}$ is characterized by the spectral density $J(\omega)=\sum_{n}g_{n}^{2}\delta(\omega-\omega_{n})$, which generally has the Ohmic-family form $J(\omega)=\eta\omega(\omega/\omega_{c})^{s-1}e^{-\omega/\omega_{c}}$ in the continuous limit of $\omega_n$. Here, $\eta$ is a dimensionless coupling constant, $\omega_{c}$ is a cutoff frequency, and $s$ is an Ohmicity index. The quantum noise is classified into sub-Ohmic when $0<s<1$, Ohmic when $s=1$, and super-Ohmic when $s>1$ \cite{RevModPhys.59.1}.

Tracing the degrees of freedom of the noise from the unitary dynamics governed by Eq. \eqref{tthmt} under the condition that the noise is initially in the vacuum state $|\{0_n\}\rangle$, we can derive an exact master equation as
\begin{equation}
\dot{\rho}(t)=-i\Omega(t)[\hat{\sigma}_+\hat{\sigma}_-,\rho(t)]+\gamma(t)\check{\mathcal L}\rho(t),\label{nmmst}
\end{equation}
where $\check{\mathcal L}\cdot= 2\hat{\sigma}_-\cdot\hat{\sigma}_+-\{\hat{\sigma}_+\hat{\sigma}_-,\cdot\}$ is the Lindblad superoperator, and $\gamma(t)\equiv -\text{Re}[\dot{x}(t)/x(t)]$ and $\Omega\equiv -\text{Im}[\dot{x}(t)/x(t)]$ are the time-dependent decay rate and renormalized frequency. The function $x(t)$ satisfies
\begin{equation}
\dot{x}(t)+i\omega_{0}x(t)+\int^{t}_{0}f(t-t')x(t')dt'=0,    \label{x(t)}
\end{equation}
under $x(0)=1$, where $f(t-t')=\int^{\infty}_{0}  J(\omega)e^{-i\omega(t-t')}d\omega$ is the correlation function of the noise. The convolution in Eq. \eqref{x(t)} renders the decoherence dynamics non-Markovian. Thus, in the presence of noise, the agent-environment interaction in each iteration is governed by Eq. \eqref{nmmst}. It is noted that $|x(t)|^2$ is the time-dependent factor of the excited-state probability of the agent. This can be seen from the solution of Eq. \eqref{nmmst} as $\langle+|\rho(t)|+\rangle=|x(t)|^2$ under the initial condition $\rho(0)=|+\rangle\langle+|$.

\begin{figure}[tbp]
\centering
\includegraphics[width=\columnwidth]{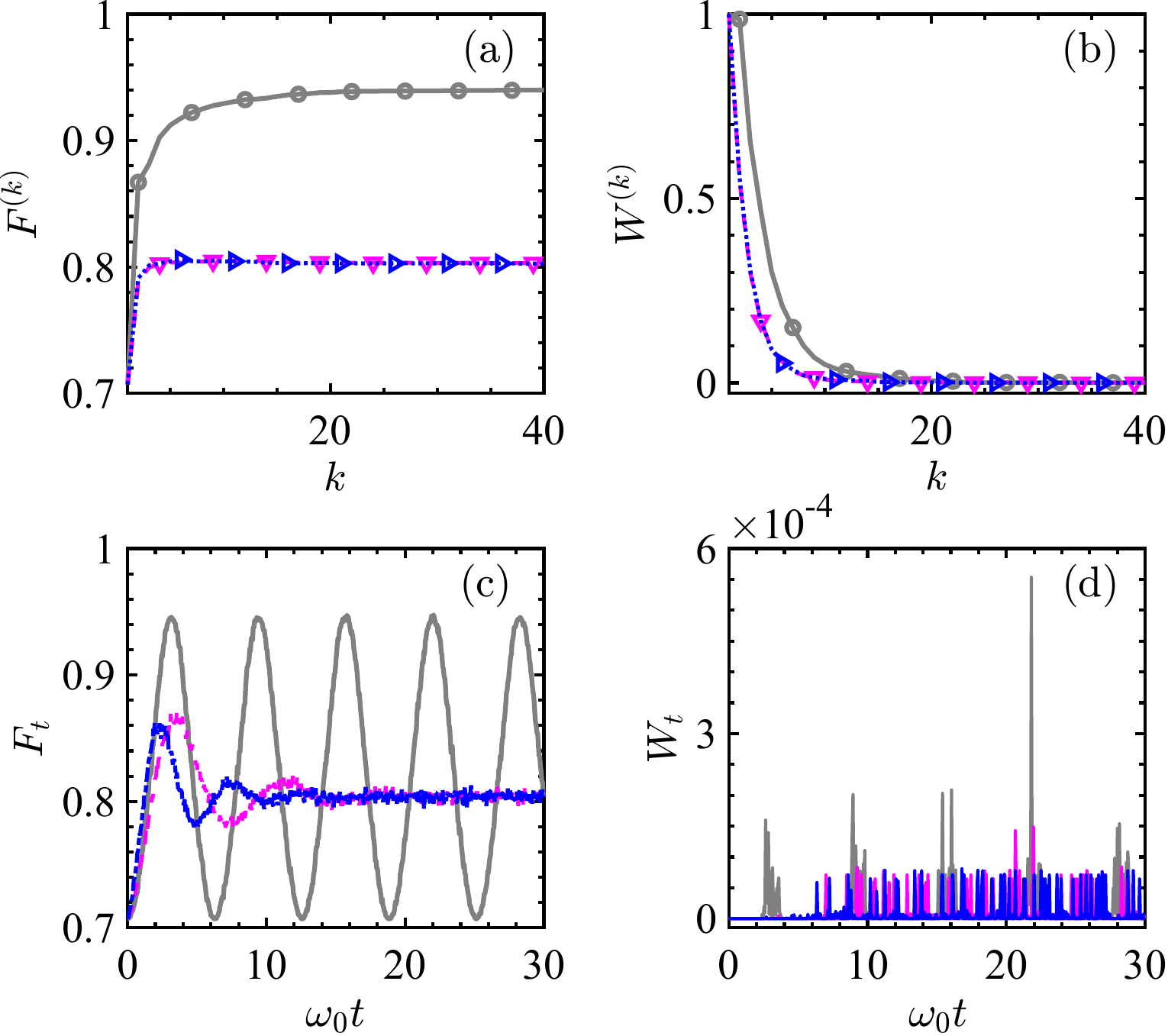}
\caption{Born-Markov approximate results. (a) Mean fidelity and (b) mean exploration parameter as functions of the iteration times $k$ for given interaction time duration $t=28\omega_0^{-1}$. (c) Mean fidelity and (d) mean exploration parameter as functions of interaction time for given iteration times $k=100$. Pink and blue lines with and without triangles represent $\omega_c/\omega_0=2$ and $20$ respectively. We use $\eta=0.1$, $s=1$, $r=0.1$, $p=1.1$, and $N=1000$. }\label{fig2}
\end{figure}

In the special case that the agent-noise coupling is weak and the time scale of $f(t-t')$ is much shorter than that of the agent, we can apply the Born-Markov approximation (BMA) to Eq. \eqref{x(t)} by replacing $x(t')$ with $x(t)$ and extending the upper limit of the integration from $t$ to $\infty$. Then we obtain $x_\text{BMA}(t)\simeq e^{-\left [ \kappa+i(\omega_{0}+\Delta(\omega_{0})) \right ]t}$, with $\kappa=\pi J(\omega_{0})$ and $\Delta(\omega_{0})=\mathcal{P} \smallint_{0}^{\infty} d \omega \left[J(\omega)/(\omega_{0}-\omega)\right]$ \cite{PhysRevE.90.022122}. This makes $\gamma(t)$ a positive constant, i.e., $\gamma_\text{BMA}(t)=\kappa$. Hence, the agent exponentially decays to its ground state and irreversibly loses its quantum coherence. The obtained mean fidelity $F^{(k)}$ for a given interaction time $t$ shows an abrupt decrease compared to the noiseless value [see Fig. \ref{fig2}(a)], although the mean exploration parameter $W^{(k)}$ still tends to zero [see Fig. \ref{fig2}(b)]. Furthermore, the evolution of $F_t$ exhibits irreversibility with $t$, although $W_t$ remains zero. This is in sharp contrast to the periodic oscillation in the noiseless case. The system parameters have little influence on this result. It means that the Born-Markov approximate decoherence degrades the fidelity, deactivates the role of the system parameters, and destroys the periodicity of the interaction time $t$ in the QRL. The result is consistent with that in Ref. \cite{PhysRevE.108.014128}.

\begin{figure}[tbp]
\centering
\includegraphics[scale=0.5]{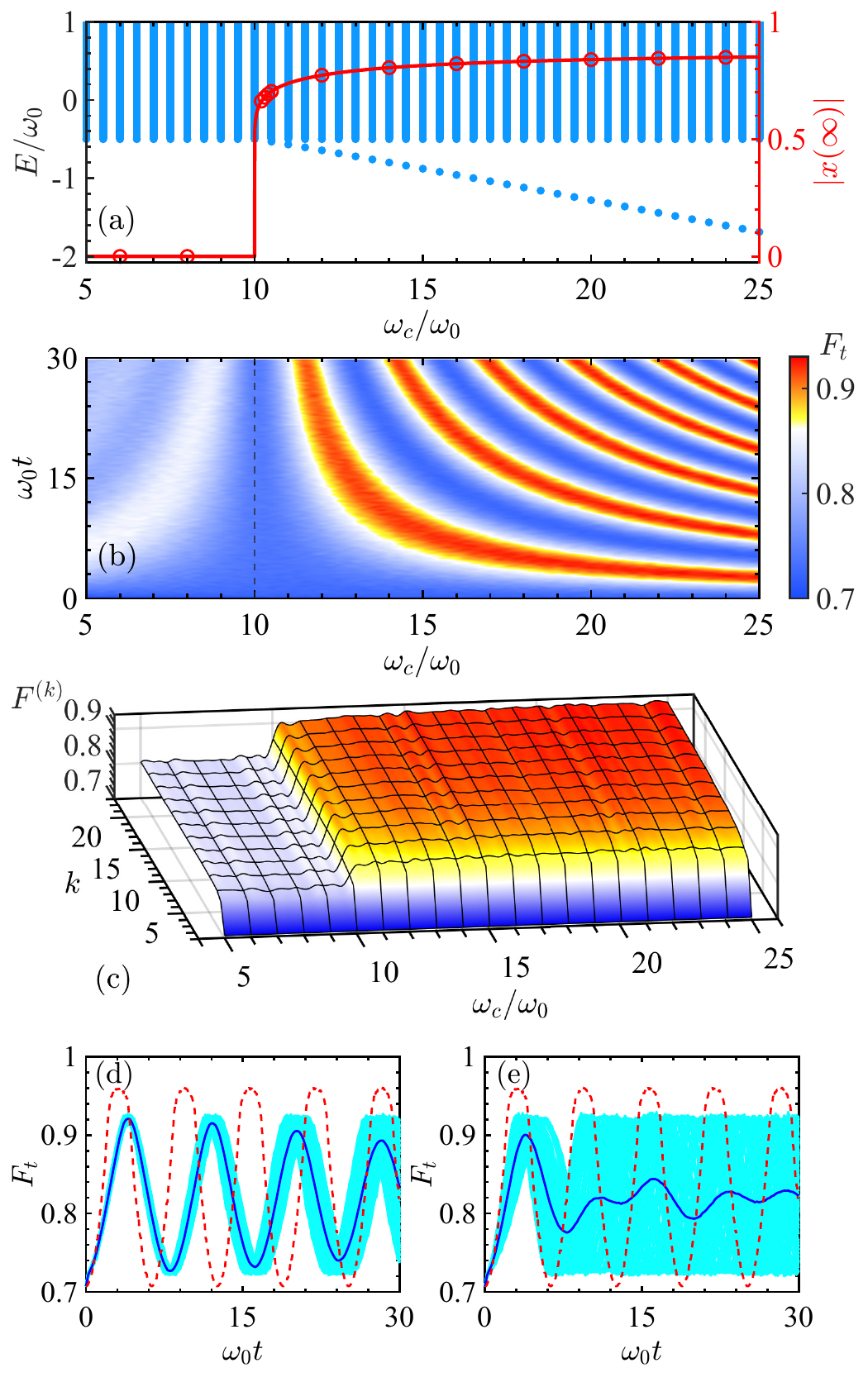}
\caption{Non-Markovian results. (a) Energy spectrum, $|x(t)|$ at $t=200\omega_0^{-1}$ (red dots), and $Z$ (red line), (b) evolution of $F_{t}$, and (c) $F^{(k)}$ as a function of the iteration times $k$ for the optimized interaction time near $t=200\omega_0^{-1}$ in different $\omega_c$. (d),(e) Plots of $F_{t}$ with dephasing in the presence of the bound state with respectively $\delta=0.1\omega_0$ in (d) and $\delta=0.6\omega_0$ in (e) for $\omega_c/\omega_0=20$. The other parameters are the same as in Fig. \ref{fig2}.}
\label{fig3}
\end{figure}
In the non-Markovian case, although Eq. \eqref{x(t)} is analytically unsolvable, its long-time form can be derived as follows. The Laplace transform $\tilde{x}(z)=\smallint_0^\infty x(t)e^{-zt}dt$ converts Eq. \eqref{x(t)} into
\begin{equation*}
    \tilde{x}(z)=\left [ z+i\omega_0+\int^{\infty}_{0}\frac{J(\omega)}{z+i\omega}d\omega \right ]^{-1}.
\end{equation*}
 Then $x(t)$ is obtained by applying the inverse Laplace transform to $\tilde{x}(z)$, which requires finding the poles of $\tilde{x}(z)$ via
\begin{equation}
Y(\bar{E})\equiv \omega_{0}-\int^{\infty}_{0}\frac{J(\omega)}{\omega-\bar{E}}d\omega=\bar{E},~(\bar{E}=iz).\label{eigE}
\end{equation}

It is interesting to find that the roots $\bar{E}$ of Eq. \eqref{eigE} are just the eigenenergies of Eq. \eqref{tthmt} up to a constant shift $\omega_0/2$ \cite{PhysRevA.87.052139}. This can be proven by expanding the eigenstate of $\hat{H}_{T}$ as $\ket{\Phi}=\mu\ket{+,\left \{ 0_{m} \right \} }+\sum_{n}\nu_{n}\ket{-,1_{n}}$. Substituting it into $\hat{H}_T\ket{\Phi}=E\ket{\Phi}$, we obtain $(E-\frac{1}{2}\omega_{0})\mu=\sum_{n}g_{n}\nu_{n}$ and $\nu_{n}=g_{n}\mu/(E+\frac{1}{2}\omega_{0}-\omega_{n})$, which result in $E-{1\over 2}\omega_0=\sum_ng_n^2/(E+{1\over2}\omega_0-\omega_n)$. This is just Eq. \eqref{eigE} in the continuous limit of $\omega_n$ by replacing $E$ with $\bar{E}-\frac{1}{2}\omega_{0}$. Since $Y(\bar{E})$ decreases monotonically in the regime of $\bar{E}< 0$, Eq. \eqref{eigE} has one isolated root denoted by $\bar{E}_b$ as long as $Y(\bar{E})<0$.

Because of the divergence of the integral in $Y(\bar{E})$ when $\bar{E}> 0$, $Y(\bar{E})$ is not analytical, and thus Eq. \eqref{eigE} has infinite roots in the regime $\bar{E}>0$, which form a continuous energy band. The eigenstates of $E_b\equiv\bar{E}_b-\frac{1}{2}\omega_{0}$ are named the bound state. With the poles of $\tilde{x}(z)$ in hand, its inverse Laplace transform is evaluated as \cite{PhysRevA.103.L010601}
\begin{equation*}
    x(t)=Ze^{-i \bar{E}_b t}+\int_{0}^{\infty} \Theta(\bar{E}) e^{-i \bar{E} t}d \bar{E},
\end{equation*}
  where
\begin{equation}
   Z=\left[1+\int_{0}^{\infty}d \omega \frac{J(\omega)}{(\bar{E}_b-\omega)^{2}} \right]^{-1},\label{zaad}
\end{equation} is from the bound state and $\Theta(\bar{E})=J(\bar{E}) /\big \{ \big[\bar{E}-\omega_{0}-\Delta(\bar{E})\big]^{2}+[\pi J(\bar{E})]^{2}\big \}$. The second term comes from the energy band and approaches zero in the long-time limit due to the out-of-phase interference. Therefore, when the bound state is absent, we have $\lim_{t \to \infty} x(t)=0$, which means a complete decoherence in the Born-Markov case. When the bound state is formed, we have $\lim_{t \to \infty} x(t)=Ze^{-i\bar{E}_bt}$, which results in $\gamma(\infty)=0$ and decoherence suppression. This reveals that the agent-environment interaction governed by $x(t)$ in the presence of noise is intrinsically determined by the feature of the energy spectrum of the total system. For the Ohmic-family spectral density, the condition of forming the bound state is evaluated from $Y(0)<0$ as $\omega_{0}< \eta\,\omega_{c}\Gamma(s)$, where $\Gamma(s)$ is the Euler gamma function.

\begin{figure}[tbp]
\centering
\includegraphics[width=\columnwidth]{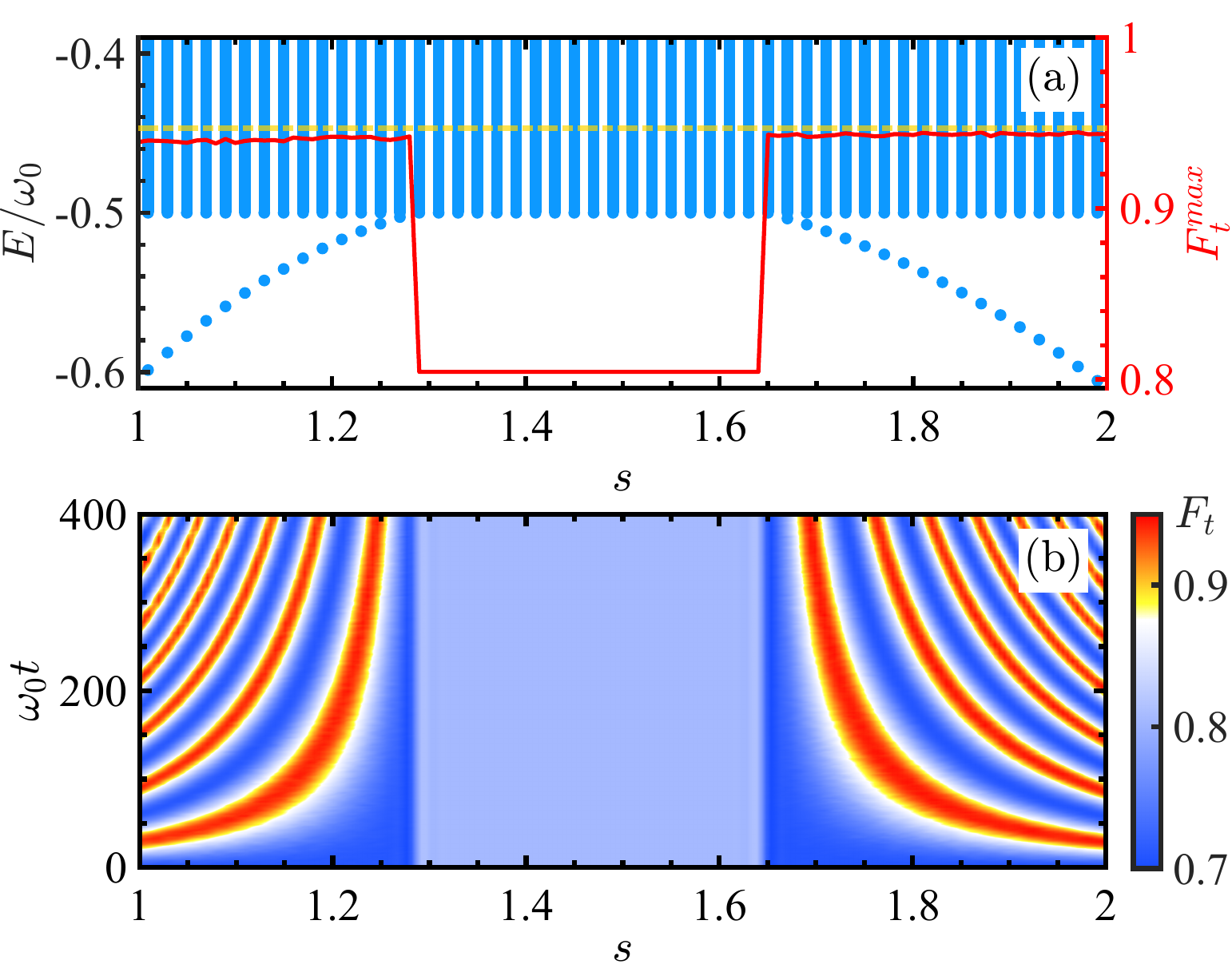}
\caption{(a) Energy spectrum and maximum mean fidelity for different $s$. (b) Evolution of $F_t$ for $k=100$ for different values of $s$. The golden dashed line in panel (a) shows the maximum mean fidelity in the ideal case. We use $\eta=0.01$,  $\omega_c=10^3/9\omega_0$, and other parameters being the same as Fig. \ref{fig2}. }
\label{fig4}
\end{figure}

The role of the bound state in restoring the QRL performance is understandable as follows. Besides solving Eq. \eqref{nmmst}, the dynamics governed by Eq. \eqref{tthmt} under $|\Psi^{(k)}(0)\rangle=\hat{D}^{(k)}|0,\{0_n\}\rangle\equiv c_+^{(k)}|+,\{0_n\}\rangle+c_-^{(k)}|-,\{0_n\}\rangle$ can be expanded in the complete basis of $|\Phi\rangle$ as
\begin{eqnarray}
|\Psi^{(k)}(t)\rangle &=&c_+^{(k)}[c_{b}e^{-iE_{b}t}|\Phi_{b}\rangle+\sum_{\alpha\in\text{Band}}c_{\alpha} e^{-iE_{\alpha} t}|\Phi_{\alpha}\rangle]\nonumber\\
&&+c_-^{(k)}|-,\{0_n\}\rangle,\label{dvd1}
\end{eqnarray}
where $c_{b/\alpha}=\langle\Phi_{b/\alpha}|+,\{0_n\}\rangle$ are the probability amplitudes of the bound state and $\alpha$th band state in $|+,\{0_n\}\rangle$. Containing the oscillating frequencies $E_{\alpha}$ continuously summed in the energy band, the contribution of the second term to $P_+(t)\equiv|\langle+,\{0_n\}|\Psi(t)\rangle|^2$ tends to zero due to the out-of-phase interference in the long-time limit.

Thus, we have $\lim_{t\rightarrow\infty}P_+(t)=|c_+^{(k)}c_b\langle+,\{0_n\}|\Phi_b\rangle|^2$. Using the definition of $|\Phi\rangle$ and the normalization condition $|\mu|^2+\sum_n|\nu_n|^2=1$, we have $c^*_b=\langle+,\{0_n\}|\Phi_b\rangle=\mu$ and $|\mu|^2=Z$ for the bound state. Therefore, we obtain
\begin{equation}
  \lim_{t\rightarrow\infty}P_+(t)=|c_+^{(k)}|^2Z^2,
\end{equation}which is consistent with the result obtained via solving Eq. \eqref{nmmst}.
When the bound state is absent, $Z=0$ and complete decoherence of the agent occurs. When the bound state is present, $Z$ tends to a value close to 1 [see Fig. \ref{fig3}(a)] and the decoherence fate of the agent is turned over. This is just the suppression of decoherence  induced by the bound state that awakens the performance of our noisy QRL.

To verify our analytical results, we plot in Fig. \ref{fig3}(a) the energy spectrum of the total system by numerically solving Eq. \eqref{eigE}. It indicates that a bound state out of the continuous band is present as long as $\omega_c>10\omega_0$ for $s=1$. Whenever the
bound state is formed, $|x(t)|$ tends to a nonzero value $Z$ [see Fig. \ref{fig3}(a)], which coincides with our analytical result. Opening a band gap, the formation of the bound state triggers a quantum phase transition. It is just this phase transition that causes the agent's abrupt jump from complete decoherence to decoherence suppression. Using the numerical result of Eq. \eqref{x(t)} in the QRL, we find that the mean fidelity $F_{t}$ saturates to $0.8$ in the absence of the bound state [see Fig. \ref{fig3}(b)], which is consistent with the Born-Markov approximate result in Fig. \ref{fig2}(c). In contrast, when the bound state is formed, $F_{t}$ not only exhibits a remarkable enhancement, but also restores its temporal periodicity in the ideal case. The convergence of $W_t$ is ensured during the calculation of $F_{t}$ in Fig. \ref{fig3}(b). By numerically optimizing $t$ in the long-time limit, in Fig. \ref{fig3}(c) we plot $F^{(k)}$ for different $\omega_c$. The clear threshold of $\omega_c$, above which $F^{(k)}$ exhibits a dramatic enhancement, matches well with the condition to form the bound state. The result verifies the restoration of the ideal performance of the noisy QRL by the formation of the bound state.

In addition to dissipative noise considered above, dephasing noise is another type of source of decoherence. To verify the robustness of our decoherence-suppression mechanism to dephasing noise, we phenomenologically model the dephasing by adding a classical noise ${1\over 2}\delta\omega_0^{\prime}(|+\rangle\langle+|-|-\rangle\langle-|)$ to Eq. \eqref{tthmt}, where $\omega_0^{\prime}$ is a random number uniformly distributed in $[-1/2,1/2]$ and $\delta$ is the noise strength. Figures \ref{fig3}(d) and \ref{fig3}(e) show the evolution of the mean values of $F_t$ obtained by averaging 100 random fluctuations and their variances for the weak and moderate dephasing noise cases, respectively. We can see from Fig. \ref{fig3}(d) that when the dephasing noise has a strength $\delta$ as high as $0.1\omega_0$, the mean value of $F_t$ and its periodicity with $t$ still show little difference from the fluctuation-free results. With the increase of the dephasing strength, $F_t$ tends to its Born-Markov approximate value and its periodicity with $t$ disappears [see Fig. \ref{fig3}(e)]. The result demonstrates the robustness of our scheme to weak dephasing noise.

Figure \ref{fig4}(a) shows the energy spectrum as a function of the Ohmicity index $s$. It confirms that the bound state is present when $\Gamma(s)>\omega_0/(\eta\omega_c)$. The presence of the bound state restores the temporal periodicity of $F_t$ [see Fig. \ref{fig4}(b)]. The maxima of $F_t$ in the presence of the bound state show tiny difference from the values in the noiseless case [see Fig. \ref{fig4}(a)]. This is substantially different from the cases without the bound state and under the Born-Markov approximation, where $F_t$ exclusively saturates to 0.8. The result verifies again that, by protecting the agent from decaying to its ground state via the formation of the bound state during the agent-environment interaction, the QRL becomes noise-resilient.

\section{Physical realization}\label{sec:realization}
Although only the Ohmic-family spectral density is considered, our result is generalizable to other forms. Inspired by a similar QRL to learn a single quantum state \cite{PhysRevA.98.042315} implemented in photonic \cite{https://doi.org/10.1002/qute.201800074} and superconducting circuit \cite{quantum2020019} systems, we here propose the following  circuit-QED platform \cite{RN10} to realize our noise-resilient scheme. The agent is modeled by a superconducting qubit. The dissipative noise is realized by an array of coupled resonators. Generally arising from elastic collisions in the dense atomic ensemble or elastic phonon scattering in a solid system \cite{book}, the dephasing noise can be neglected in our circuit-QED setup. Besides, given the fact that the circuit QED system operates at a low temperature, the finite temperature effect can also be neglected.

\begin{figure}[tbp]
\centering
\includegraphics[width=\columnwidth]{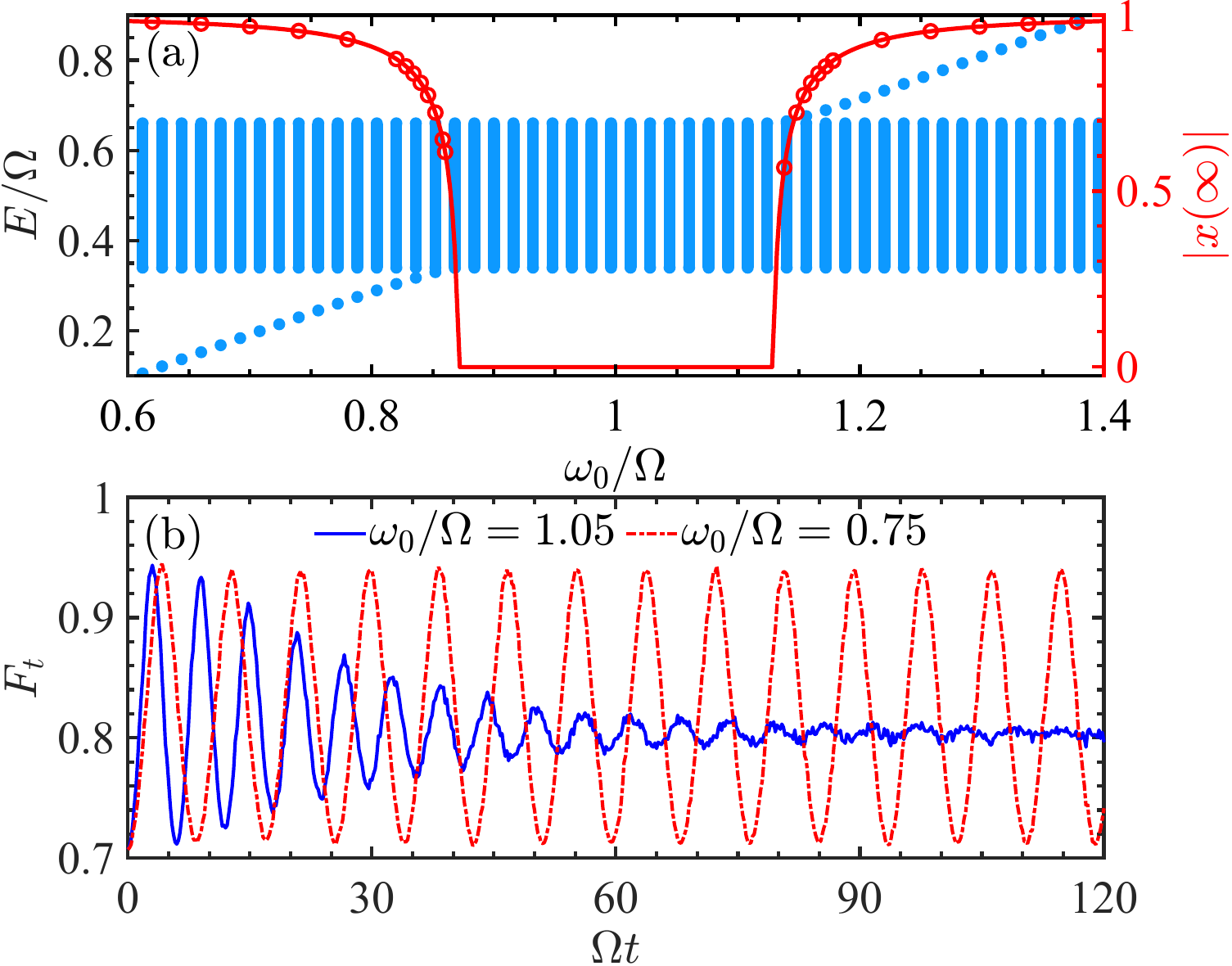}
\caption{(a) Energy spectrum of a circuit QED qubit interacting with an array of coupled resonators for different $\omega_0$. The red line is obtained by analytically calculating $Z$ and the red dots are derived numerically from the long-time solution of Eq. \eqref{x(t)}. (b) Mean fidelity $F_t$ as a function of the interaction time $t$ with and without the bound state. We use $g=0.02\Omega$ and $\zeta=0.08\Omega$ and the other parameters are the same as in Fig. \ref{fig3}.}\label{fig5}
\end{figure}

The dissipative-noise and interaction Hamiltonians are \begin{eqnarray}
    \hat{H}_{b}&=&\sum_{j=1}^{N}[\Omega\hat{b}_{j}^{\dagger}\hat{b}_{j}+\zeta(\hat{b}_{j}^{\dagger}\hat{b}_{j+1}+\hat{b}_{j+1}^{\dagger}\hat{b}_{j})],\\
    \hat{H}_i&=&g(\hat{\sigma}_-\hat{b}^{\dagger}_1+\text{H.c.}).
\end{eqnarray}
After applying the Fourier transform on $\hat{b}_j$, we have $\hat{H}_b=\sum\omega_k\hat{b}_k^{\dagger}\hat{b}_k$ with the dispersion relation $\omega_k=\Omega+2\zeta\cos k$. The spectral density is $J(\omega)=g^{2}(2\pi\zeta^{2})^{-1}\sqrt{4\zeta^{2}-(\omega-\Omega)^{2}}$, which takes the form of a finite bandgap structure. Figure \ref{fig5}(a) shows the energy spectrum of the total system formed by the qubit and the coupled resonator array for different values of $\omega_0$. The bound state is present when $\omega_0>1.14\Omega$ and $\omega_0<0.86\Omega$. The long-time behavior of $|x(t)|$ obtained by numerically solving Eq. \eqref{x(t)} is exactly the same as $Z$ analytically evaluated from Eq. \eqref{zaad}. From Fig. \ref{fig5}(b), we can see the enhancement of the mean fidelity in the presence of the bound state in comparison with the situation without the bound state, which demonstrates the effectiveness of our noise-resilient scheme.

Various methods have been proposed to manipulate the spectral density \cite{PhysRevLett.97.016802,PhysRevA.78.010101}. The bound state and its dynamical effect have been systematically revealed \cite{PhysRevLett.109.170402,PhysRevLett.132.090401,PhysRevLett.131.050801} and experimentally observed in circuit QED \cite{RN10} and ultracold atom \cite{RN11,RN14} systems. The QRL \cite{Albarrán-Arriagada_2020} has been implemented on an IBM quantum computer \cite{RN6}, whose advantage over variational quantum eigensolver algorithms in saving resources has been demonstrated. A similar QRL to learn a single quantum state \cite{PhysRevA.98.042315} has been implemented in photonic \cite{https://doi.org/10.1002/qute.201800074} and superconducting circuit \cite{quantum2020019} systems. All this progress provides essential support for the experimental realization of our findings. Although only the effect of  noisy on the QRL is studied, our result is hopefully applicable in the noise suppression of other quantum machine learning.
\section{Conclusion}\label{sec:conclusion}
In summary, we have proposed a noise-resilient QRL scheme for the quantum eigensolver of a two-level system. We have discovered a mechanism to overcome the detrimental impact of noise-induced non-Markovian decoherence on the QRL. It has been revealed that, accompanying the formation of a bound state in the energy spectrum of the total system consisting of the agent and quantum noise, the mean fidelity approaches its ideal value and oscillates periodically with the interaction time in the same way as the ideal behavior. It is important to note that our work aims to highlight the underlying noise-resilient mechanism, rather than presenting an efficient algorithm. The central contribution lies in uncovering this general mechanism, which is not necessarily confined to single-qubit models, thereby offering new guidance for designing noise-resilient NISQ algorithms. Efficiently eliminating the destructive influence of noise, our result provides a guideline for the realization of QRL in the NISQ era.

\begin{acknowledgments}
This work is supported by the National Natural Science Foundation of China (Grants No. 12275109, No. 92576202, and No. 12247101), the Quantum Science and Technology-National Science and Technology Major Project (Grant No. 2023ZD0300904), the Fundamental Research Funds for the Central Universities (Grant No. lzujbky-2025-jdzx07), and the Natural Science Foundation of Gansu Province (Grants No. 22JR5RA389 and No. 25JRRA799).
\end{acknowledgments}

\bibliography{reference}

\end{document}